\documentclass[a4paper,11pt]{article}
\usepackage{amsmath,amssymb}
\usepackage{jinstpub} 

\title{\boldmath The nylon balloon for xenon loaded liquid scintillator in KamLAND-Zen 800 neutrinoless double-beta decay search experiment}

\author[a,1]{Y.~Gando,\note{Corresponding author.}}
\author[a]{A.~Gando,}
\author[a]{T.~Hachiya,}
\author[a,2]{S.~Hayashida,\note{Present address: Imperial College London, Department of Physics, Blackett Laboratory, London SW7 2AZ, UK.}}
\author[a]{K.~Hosokawa,}
\author[a]{H.~Ikeda,}
\author[a]{T.~Mitsui,}
\author[a]{T.~Nakada,}
\author[e]{S.~Obara,}
\author[a,f]{H.~Ozaki,}
\author[a]{J.~Shirai,}
\author[a,3]{K.~Ueshima,\note{Present address: National Institutes for Quantum and Radiological Science and Technology (QST), Hyogo 679-5148, Japan.}}
\author[a]{H.~Watanabe,}

\author[a]{S.~Abe,}
\author[a]{K.~Hata,}
\author[a]{A.~Hayashi,}
\author[a]{Y.~Honda,}
\author[a]{S.~Ieki,}
\author[a,b]{K.~Inoue,}
\author[a]{K.~Ishidoshiro,}
\author[a]{S.~Ishikawa,}
\author[a]{Y.~Kamei,}
\author[a]{K.~Kamizawa,}
\author[a]{Y.~Karino,}
\author[a]{N.~Kawada,}
\author[a]{T.~Kinoshita,}
\author[a,b]{M.~Koga,}
\author[a]{S.~Matsuda,}
\author[a]{H.~Miyake,}
\author[a,b]{K.~Nakamura,}
\author[a]{K.~Nemoto,}
\author[a]{A.~Ono,}
\author[a]{N.~Ota,}
\author[a]{S.~Otsuka,}
\author[a]{Y.~Shibukawa,}
\author[a]{I.~Shimizu,}
\author[a]{Y.~Shirahata,}
\author[a]{K.~Soma,}
\author[a]{A.~Suzuki,}
\author[a]{A.~A.~Suzuki,}
\author[a]{T.~Takai,}
\author[a]{A.~Takeuchi,}
\author[a]{K.~Tamae,}
\author[a]{Y.~Teraoka,}
\author[a]{Y.~Wada,}

\author[b,4]{D.~Chernyak,\note{Present address: Department of Physics and Astronomy, University of Alabama, Tuscaloosa, AL 35487, USA}}
\author[b,5]{A.~Kozlov,\note{Present address: National Research Nuclear University  ``MEPhI'' (Moscow Engineering Physics Institute), Moscow, 115409, Russia}}

\author[c]{S.~Yoshida,}

\author[d]{S.~Umehara,}
\author[b,d,6]{Y.~Takemoto,\note{Present address: Kamioka Observatory, Institute for Cosmic Ray Research, The University of Tokyo, Hida, Gifu 506-1205, Japan}}

\author[g]{K.~Fushimi,}
\author[g]{S.~Hirata,}

\author[h]{C.~Grant,}
\author[h]{A.~Li,}

\author[i]{J.G.~Learned,}
\author[i]{J.~Maricic,}

\author[b,j]{B.E.~Berger,}
\author[b,j]{B.K.~Fujikawa,}

\author[k]{S.~Fraker,}
\author[k]{A.~Herman,}
\author[k]{E.~Krupczak,}
\author[k]{G.~L.~Pease,}
\author[k]{L.A.~Winslow,}

\author[b,l]{Y.~Efremenko,}

\author[m]{H.J.~Karwowski,}
\author[m]{D.M.~Markoff,}
\author[m]{W.~Tornow,}

\author[n]{T.~O'Donnell,}
\author[n]{S.~Dell'Oro,}

\author[b,o]{J.A.~Detwiler,}
\author[b,o]{S.~Enomoto,}

\author[b,p]{M.P.~Decowski}

\affiliation[a]{Research Center for Neutrino Science, Tohoku University, Sendai 980-8578, Japan}
\affiliation[b]{Kavli Institute for the Physics and Mathematics of the Universe (WPI), The University of Tokyo Institutes for Advanced Study, The University of Tokyo, Kashiwa, Chiba 277-8583, Japan}
\affiliation[c]{Graduate School of Science, Osaka University, Toyonaka, Osaka 560-0043, Japan}
\affiliation[d]{Research Center for Nuclear Physics, Osaka University, 1-10 Mihogaoka, Ibaraki city, Osaka 567-0042, Japan}
\affiliation[e]{Frontier Research Institute for Interdisciplinary Sciences, Tohoku University, Sendai 980-8578, Japan}
\affiliation[f]{Graduate Program on Physics for the Universe (GP-PU), Tohoku University, Sendai 980-8578, Japan}
\affiliation[g]{Faculty of Integrated Arts and Science, University of Tokushima, Tokushima, 770-8502, Japan}
\affiliation[h]{Department of Physics, Boston University, Boston, Massachusetts 02215, USA}
\affiliation[i]{Department of Physics and Astronomy, University of Hawaii at Manoa, Honolulu, Hawaii 96822, USA}
\affiliation[j]{Lawrence Berkeley National Laboratory, Berkeley, California 94720, USA}
\affiliation[k]{Massachusetts Institute of Technology, Cambridge, Massachusetts 02139, USA}
\affiliation[l]{Department of Physics and Astronomy, University of Tennessee, Knoxville, Tennessee 37996, USA}
\affiliation[m]{Triangle Universities Nuclear Laboratory, Durham, North Carolina 27708, USA and Physics Departments at Duke University, North Carolina Central University, and the University of North Carolina at Chapel Hill}
\affiliation[n]{Center for Neutrino Physics, Virginia Polytechnic Institute and State University, Blacksburg, Virginia 24061, USA}
\affiliation[o]{Center for Experimental Nuclear Physics and Astrophysics, University of Washington, Seattle, Washington 98195, USA}
\affiliation[p]{Nikhef and the University of Amsterdam, Science Park, 1098XG, Amsterdam, the Netherlands}

\emailAdd{gando@awa.tohoku.ac.jp}

\abstract{
The KamLAND-Zen 800 
experiment is searching for the neutrinoless
double-beta decay of $^{\rm 136}$Xe by using $^{\rm 136}$Xe-loaded liquid scintillator.
The liquid scintillator is enclosed inside a balloon made of thin,
transparent, low-radioactivity film that we call Inner Balloon (IB).
The IB, apart from guaranteeing the liquid containment, also allows to
minimize the background from cosmogenic muon-spallation products 
and $^{\rm 8}$B solar neutrinos.
Indeed these events could contribute to the total counts in the region of
interest around the Q-value of the double-beta decay of $^{\rm 136}$Xe.
In this paper, we present an overview of the IB and describe the various
steps of its commissioning minimizing the radioactive contaminations, 
from the material selection,
to the fabrication of the balloon and its installation
inside the KamLAND detector. Finally, we show the impact of the IB on the
KamLAND background as measured by the KamLAND detector itself.
}


\arxivnumber{2104.10452} 

\collaboration{
 KamLAND-Zen collaboration}

\begin{document}
\maketitle
\flushbottom

\section{Introduction}

Neutrinoless double-beta ($0\nu\beta\beta$) decay searches are
direct probes of Majorana neutrinos and lepton number
violation~\cite{Majorana}.  Many Grand Unified Theories predict a 
Majorana nature for the neutrino, and connect the neutrino's 
small mass to grand unification scale physics through the see-saw
mechanism~\cite{Minkowski,SeeSaw1,SeeSaw2}.  In addition, the matter-antimatter asymmetry
of the universe could be explained by some leptogenesis models, which also predict
Majorana neutrinos~\cite{Leptogenesis}. 
Regardless of theoretical motivations, the detection of $0\nu\beta\beta$ decay 
would signify lepton number violation, and would mark the first observation 
of  asymmetric matter creation in the laboratory. 
For these reasons, searches for $0\nu\beta\beta$ decay are being 
vigorously pursued around the world. 

KamLAND-Zen is searching for $0\nu\beta\beta$ decay 
of the nucleus $^{136}$Xe. 
Experimentally, $0\nu\beta\beta$ decay searches require a large
nuclei mass and a low-background environment 
given the very long half-lives expected for $0\nu\beta\beta$ 
decay (>10$^{\rm 26}$ yr) and the \textit{Q}-values of a few MeV.  
In KamLAND-Zen, the energy region surrounding 
the \textit{Q}-value of the $^{\rm 136}$Xe decay, 
2.458~MeV~\cite{Qvalue}, is subject to environmental 
background from contamination present in
the existing KamLAND detector material, 
including $^{\rm 238}$U-series and $^{\rm 232}$Th-series decays.
To minimize such contamination, we developed a low-background 
nylon balloon to contain xenon-loaded liquid scintillator (Xe-LS) 
for the KamLAND-Zen experiment. 
The first phase, KamLAND-Zen 400 using a clean nylon balloon and 
enriched xenon up to 383~kg started in 2011 and obtained 
a lower limit for the $0\nu\beta\beta$ decay half-life of 
$T^{0\nu}_{1/2} > 1.07 \times 10^{26}$ yr at 90~\% C.L.~\cite{zen400final}. 
The next phase of KamLAND-Zen, KamLAND-Zen 800 with a cleaner 
balloon which was made with improved methods 
and new techniques started in 2018. 
In this paper,
we describe the design of the balloon
(Section~\ref{sec:innerballoon}),
the component materials
(Section~\ref{sec:filmselection}),
the production methods and clean environment procedures
(Section~\ref{sec:fabrication}),
the installation of the balloon into the KamLAND detector
(Section~\ref{sec:installation}), 
and finally the contamination analysis of the balloon 
(Section~\ref{sec:analysis}).

\section{Detector} \label{sec:detector}
\subsection{KamLAND detector}

\begin{figure}[htb]
\centering
\includegraphics[height=3.0in]{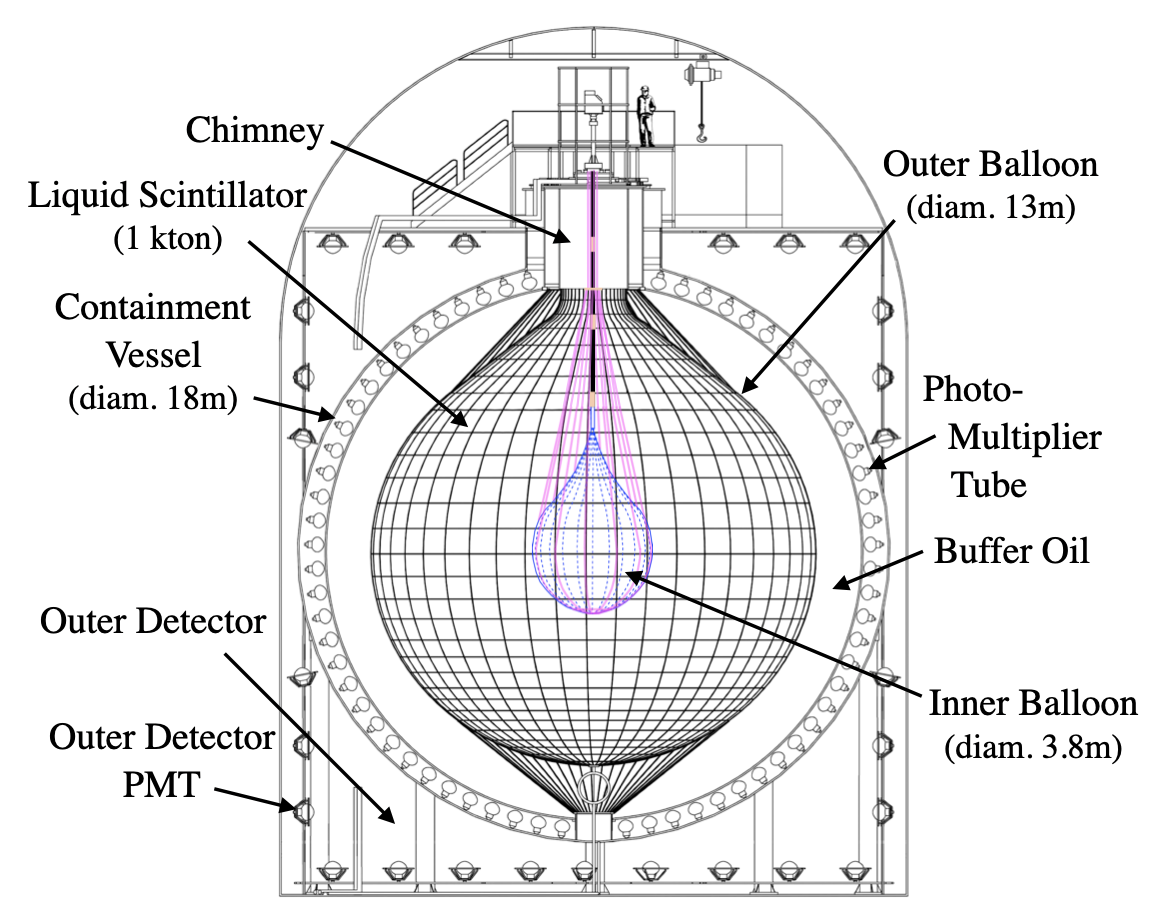}
\caption{Schematic view of KamLAND with the IB for KamLAND-Zen 800.}
\label{fig:KamLAND}
\end{figure}

KamLAND (Kamioka Liquid scintillator Anti-Neutrino Detector) is a
neutrino detector based on 1,000~tons of purified liquid-scintillator
(LS) in a low background environment with 2,700~m.w.e. 
rock overburden in the Kamioka mine in Japan~\cite{KamLAND}.  
The collaboration 
has published measurements of reactor anti-neutrinos~\cite{KL_reactor},
geological anti-neutrinos~\cite{KL_geonu_nature,KL_geonu_naturegeoscience},
and solar neutrinos~\cite{KL_solar_b8,KL_solar}.
Figure~\ref{fig:KamLAND} shows a schematic view of the KamLAND
detector.  It consists of a spherical stainless-steel tank,
photomultiplier tubes (PMTs), buffer oil, and a 13-m-diameter balloon
(outer balloon, OB) filled with LS (KamLAND LS).  
The KamLAND-Zen experiment introduces a second smaller balloon 
(inner balloon, IB) 
of LS loaded with $^{136}$Xe-enriched xenon at the center of the detector. 

The outer balloon is made of five-layered nylon/EVOH film and is 135~$\mu$m
thick.  The buffer oil between the outer balloon and the
stainless-steel vessel shields the LS against external radiation.  
Isotropic emission photons from ionizing radiation in the LS 
are detected by an array of 1,325 17-inch PMTs and 554 20-inch PMTs
covering 34\% of the inner surface of the detector.  
The vertex position and energy of the physics events are reconstructed from the timing
correlation and charge of the hit PMTs, accounting for the transparency
of the LS, the balloon film and buffer oil.  
The achieved resolution is $\sim$12~cm/$\sqrt{\rm E(MeV)}$ 
for the vertex reconstruction and $\sim$6.4\%/$\sqrt{\rm E(MeV)}$ 
for the energy~\cite{KL_reactor}.
The ultra-low background ($\sim$10$^{\rm -17}$~g/g for uranium and thorium~\cite{KL_solar}) 
and huge volume (1,200~m$^{\rm 3}$) of the LS make KamLAND one of the most sensitive
LS detector environments for $0\nu\beta\beta$ decay searches. 

\subsection{KamLAND-Zen 800}

KamLAND-Zen (KamLAND Zero neutrino double beta decay 
search)~\cite{KL-Zen} is an experiment searching for the 
$0\nu\beta\beta$ decay of $^{\rm 136}$Xe inside KamLAND, 
where the IB filled with Xe-LS is 
suspended at the center of the detector.  
KamLAND-Zen 800 uses 745~kg
xenon with a 91\% isotopic abundance of $^{\rm 136}$Xe.  Xenon is
soluble in LS up to 3.1\% by weight at atmospheric pressure.  It
can be extracted from LS by vacuum degassing and nitrogen bubbling.
Backgrounds 
from radioactive elements 
in the Xe-LS are reduced by purifying the LS~\cite{KL_solar}
and xenon~\cite{XMASS_purification} separately.  The Xe-LS is
transparent at its peak emission wavelength of $\sim$385~nm.

The Q-value of $^{\rm 136}$Xe $\beta\beta$ decay is 2.458~MeV.
Several backgrounds exist in this energy region: environmental
radiation from $^{\rm 238}$U-series, pileup events from 
$^{\rm 232}$Th-series in LS, 
cosmogenic muon spallation products
of carbon ($^{\rm 10}$C, etc.) and xenon, 
and solar $^{\rm 8}$B neutrinos.
The $^{\rm 10}$C and pileup events can be rejected analytically: 
the $^{\rm 10}$C is vetoed through the
triple coincidence of muon, neutron, and beta/gamma-ray 
events from $^{\rm 10}$C decay with a tagging 
efficiency estimated at about~90\%~\cite{KL_spallation}. 
The pileup occurs in close time coincidence within 
a single event window ($\sim$200~ns) and is also a possible background, 
but it can be rejected using long-time sequential 
decays as long as rates from other backgrounds such as 
$^{\rm 210}$Po are low~\cite{LongTimeTagging}. 
To reduce backgrounds proportional
to the LS volume, such as spallation products and solar $^{\rm 8}$B
neutrinos, it is advantageous to maximally load the $\beta\beta$ nuclei
in the smaller volume of the inner balloon rather than using the
full LS volume of KamLAND at a lower loading percentage.  
The balloon itself introduces radioactive
backgrounds proportional to its total weight, therefore low-radioactivity
materials with minimal mass are required.  

Due to its spherical shape, the access 
space into the detector is limited to a 50-cm-diameter 
opening at the top of KamLAND.  
The IB is therefore installed and suspended from the
chimney region at the top of the detector.  
To meet these mechanical constraints as well as to achieve 
the background requirements,
we selected a balloon vessel made of thin, 
transparent, and foldable low-background polymeric film for the
$0\nu\beta\beta$ decay search in KamLAND-Zen.

\section{Inner Balloon} \label{sec:innerballoon}

\subsection{Inner balloon design}

\begin{figure}[htb]
\centering
\includegraphics[height=3.5in]{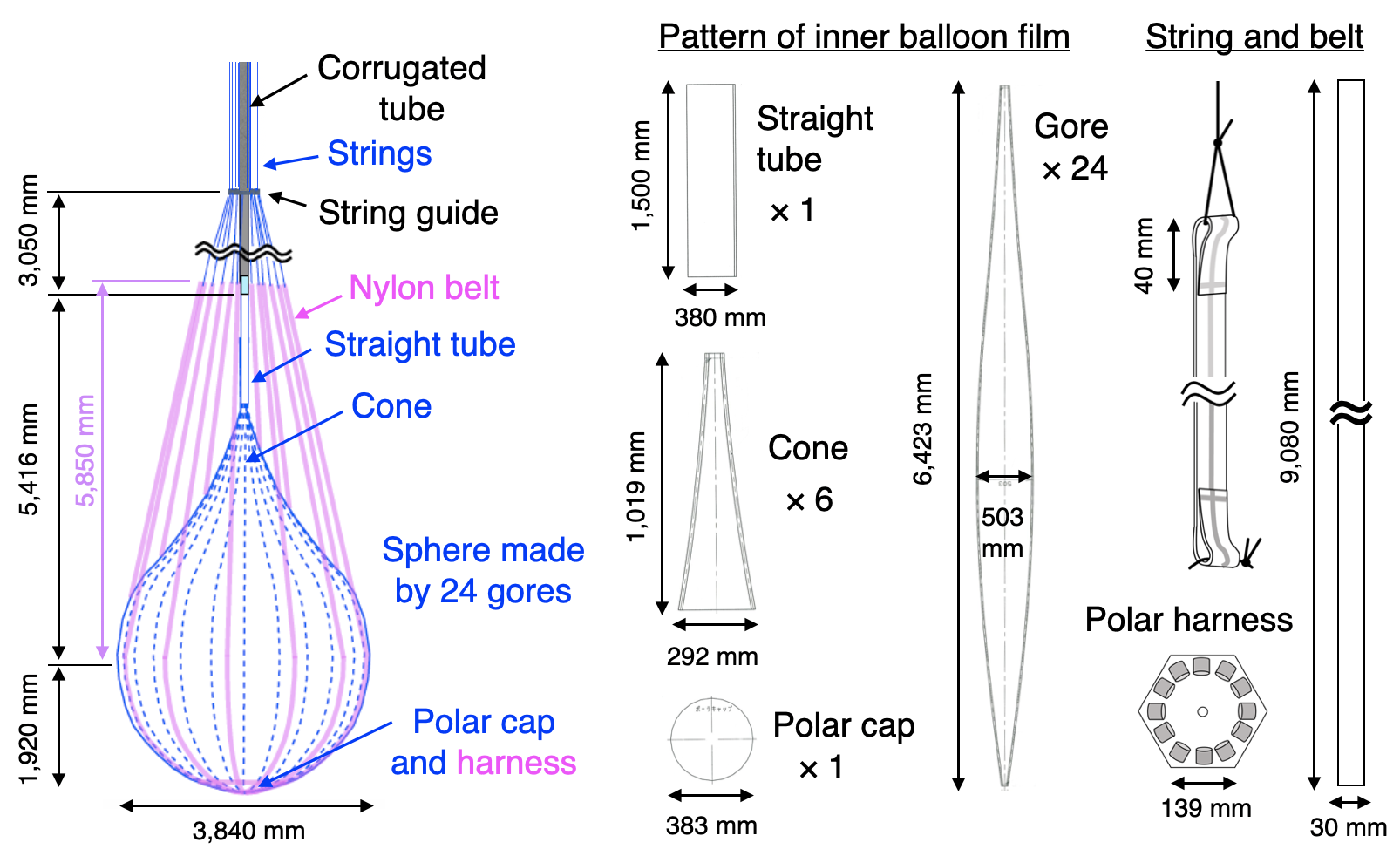}
\caption{Schematic view of the IB, related structures,
        and the IB film patterns.  The IB is
        suspended by a system of nylon belts and a harness (magenta) which support the total weight of 
        the Xe-LS and balloon.  The belts are connected to Vectran
        strings attached to load cells, and the weight supported
        by each string is monitored.}
\label{fig:mini-balloon}
\end{figure}

The IB is designed in a teardrop shape with smooth curves
to avoid stress concentration, as shown in Figure~\ref{fig:mini-balloon}.
We designed the IB with a volume of 31.5m$^{3}$ determined by 
the amount of xenon of 750 kg and the solubility to LS of 3.0 wt\%, 
which leads to the balloon radius to be 1,920~mm.
The IB consists of a main sphere formed from 24 gores,
a lower polar cap, an upper cone formed from six film segments, and
a 1,500-mm-long straight tube.  
The 25~$\mu$m-thick nylon film sections are connected by
the heat-welding method described in Section~\ref{sec:fabrication}.
The sections are overlaid by 15~mm for welding at the seams.  

The IB is supported by twelve 30-mm-wide nylon belts 
made of the same material as the IB itself.
The connection of these belts to the polar harness at 
the bottom of the balloon is described in Section~\ref{sec:HangingBelt}.
The belts are connected to twelve 0.4-mm-diameter strings
(Fujinoline Corporation) made of Vectran (Kuraray Corporation) 
in the region of the corrugated tube, well above the IB body. 
The radioactive contamination
of these strings is not sufficiently small for them to be employed
near the IB.  The strings enable easier handling of the
balloon during installation.  The top ends of the strings are connected
to a rotary ratchet that can adjust their lengths in 1.5-mm steps.
The ratchet is suspended by a load cell (SRM200-50N, Controls Japan
Company) to monitor the total weight of the IB.  A
string guide made of polyether ether ketone (PEEK) leads the strings
into the chimney area of KamLAND.  To prevent expansion or contraction
due to head pressure we use a corrugated tube for the connection
between the straight tube at the top of the IB and the
KamLAND chimney (Section~\ref{sec:corrugatedtube}).

\subsection{Xenon-loaded liquid scintillator}
\label{sec:XenonLS}

To avoid damage to the IB, the density of Xe-LS is tuned
to very closely match the KamLAND LS in the outer balloon.  The 
KamLAND LS density is 0.77728~kg/L 
at 15 $^\circ$C.  It consists
of 20\% pseudocumene (1,2,4-trimethylbenzene, 0.8797~kg/L), 80\%
dodecane (CH$_{\rm 3}$(CH$_{\rm 2}$)$_{\rm 10}$CH$_{\rm 3}$, 0.7526~kg/L), and 1.36~g/L of PPO (2,5-diphenyloxazole).
The density increase caused by the xenon dissolved in the LS could be
counterbalanced by reducing the pseudocumene/dodecane fraction, but
this would reduce the light yield, which depends on the amount of pseudocumene.  
We therefore replace the dodecane with decane 
(CH$_{\rm 3}$(CH$_{\rm 2}$)$_{\rm 8}$CH$_{\rm 3}$, 0.7339~kg/L) 
to maximize the pseudocumene fraction in
xenon-loaded LS.  The decane-based LS contains
18\% pseudocumene, 82\% decane, and 2.38~g/L of PPO.  
The density of the Xe-LS can be controlled with an accuracy of 
0.02\% by tuning the pseudocumene/decane fraction.  
The actual density of the Xe-LS is slightly heavier than
the outer KamLAND LS.  
It is controlled within 0.1\% to the KamLAND LS at all times, 
from installation through data acquisition, to avoid 
damage to the IB caused by deformation 
if this was to float in the KamLAND LS.  
The final density difference between the outer KamLAND LS and
inner Xe-LS is about 0.02\%.
The isotopic abundance of $^{\rm 136}$Xe is 90.9\%, 
and the remaining components are $^{\rm 124}$Xe--$^{\rm 134}$Xe. 
The total amount of $^{\rm 136}$Xe in the IB is 
677~kg.

\subsection{Corrugated tube} \label{sec:corrugatedtube}

The uppermost straight section of the inner vessel consists of a hard nylon 
tube.  This design avoids contraction or expansion at the surface
of LS due to the head pressure between the outer KamLAND LS and the
inner Xe-LS.  This tube has to be bent during the IB installation, 
so we used a corrugated nylon tube (CYLG-95B, PMA Company).  
The $^{\rm 238}$U contamination of the corrugated
tube was measured by ICP-MS at NTT-AT Co.~to be 
9$\times$10$^{\rm -12}$~g/g after washing, a level
acceptable for use in KamLAND. 
However, this contamination level is about five times 
larger than that of the IB film. 
As the tube thickness is 1~mm, 40 times thicker than the film, the total
$^{\rm 238}$U contamination is 200 times that of the film, per unit surface area. 
To avoid $\gamma$-rays from the corrugated tube introducing backgrounds 
into the Xe-LS in the spherical area of the IB, the lower
straight section is made of IB film for the first 1.5~m
above the body of the IB.  

\begin{figure}[htb]
\centering
\includegraphics[height=2.3in]{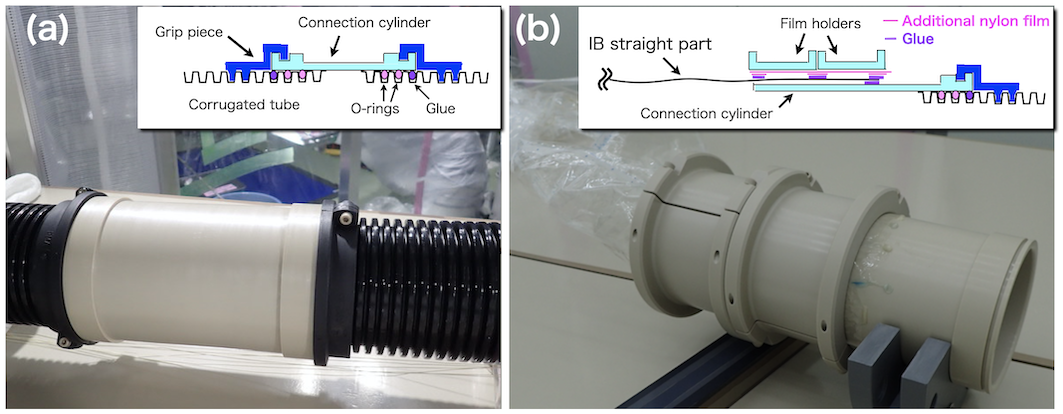}
\caption{(a) Picture and schematic view of the connection between
        corrugated tubes via a connection cylinder.
        (b) Connection between the straight tube of the IB
        and the connection cylinder.
        The additional 25~$\mu$m-thick nylon film 
        between the film holders and the IB film avoids IB damage 
        from the film holders.
        }
\label{fig:CorrugatedTube}
\end{figure}

The total length of corrugated tube is about 6~m 
and the diameter is about 10~cm.  
It is separated into three sections to allow
cleaning in the clean-room and to allow it to be kept straight 
during transportation.  
The three sections are connected with a 
custom-made cylinder made of PEEK
as commercial connection cylinders were found not to 
be compatible with our LS.  This connection uses fluoro-rubber
O-rings, which can be used in LS. 
A grip piece is used to secure the connection between the 
cylinder and the tube and prevent it from falling out. 
Figure~\ref{fig:CorrugatedTube}(a) shows the connection between the
cylinder and corrugated tubes.  The connection between the corrugated
tube and the straight film tube of the IB is also via a
PEEK cylinder, as shown in Figure~\ref{fig:CorrugatedTube}(b).  The
connection between the film tube and the rest of the IB
is described in Section~\ref{sec:FilmWithCylinder}.

\section{Film Selection} \label{sec:filmselection}

We selected a custom-made nylon film for the IB because
no commercial film satisfied our requirements
on the chemical compatibility and radioactive contamination. 
These are described in the present section, together with the measured values.

\subsection{Chemical compatibility and handling}

Only a few kinds of film materials are compatible with a pseudocumene-based
LS: ethylene tetrafluoroethylene (ETFE), nylon,
and ethylene-vinylalcohol copolymer (EVOH).  
The outer balloon was fabricated using nylon and EVOH. 
However, the multi-layered film was made with glue 
which has radioactive impurities. 
We therefore had to select a single-layer film.
We assessed chemical compatibility
by checking for a decrease in the transparency of LS after soaking
the films.  For nylon and EVOH films, the transparency of 
the liquid samples after the film soaking 
was consistent within measurement error to a reference
sample of the same scintillator.

EVOH has high gas impermeability. 
However, an EVOH film was easily torn during the production of 
an 80-cm-diameter test balloon.  
ETFE showed creep under tension, so an ETFE vessel could not stably 
maintain its shape.  
A nylon film was therefore the only candidate for the IB film.

\subsection{Radioactive contamination}

\begin{figure}[htb]
\centering
\includegraphics[height=1.7in]{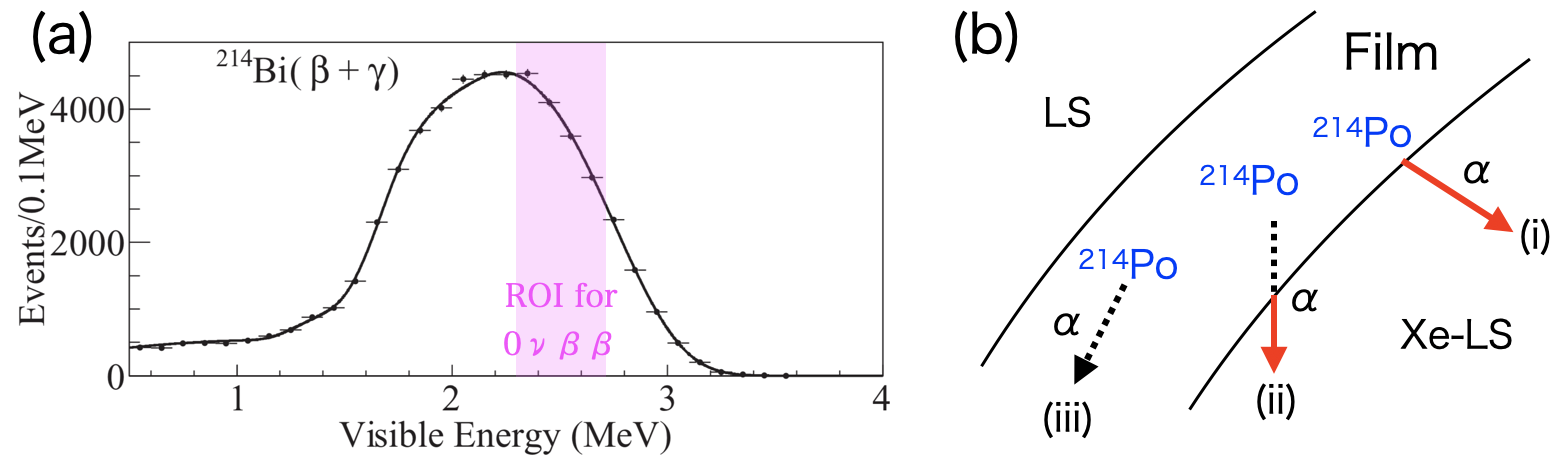}
\caption{(a) The energy spectrum of $^{214}$Bi($\beta$+$\gamma$) 
        with the $0\nu\beta\beta$ decay region of interest (ROI) 
        in KamLAND-Zen~\cite{KL-Zen}. 
        (b) Light output from $\alpha$-rays depending on the 
        penetration route in the film.
        (i)~Light output in Xe-LS (or LS) can be detected.
        (ii)~The detection probability depends on the energy deposit
        in Xe-LS (or LS).
        (iii)~There is no scintillation light; the event can not be
        detected.}
\label{fig:AlphaInFilm}
\end{figure}

The $^{\rm 214}$Bi in the $^{\rm 238}$U-series is one of the primary
backgrounds for our $0\nu\beta\beta$ decay search 
(Figure~\ref{fig:AlphaInFilm}(a)).
The target level of $^{\rm 238}$U in the balloon film 
is 10$^{\rm -12}$~g/g for a film thickness of 25~$\mu$m assuming the secular equilibrium. 
This level was set at 
$\mathcal O$(10\%) of the estimated event rate from $^{\rm 10}$C
that survives a $^{\rm 10}$C tagging veto.  
Commercial nylon films have high impurity levels
($\sim$10$^{\rm -9} $~g/g at $^{\rm 238}$U).  However, cleaner films
can be specially made without ``filler''.  The filler is inorganic
lubricant used to avoid static electricity and to prevent the film
sticking to itself.  We used custom-made Nylon-6 film produced
by Toyobo Company from pellets produced by Ube Industries.
The $^{\rm 238}$U level in this nylon film was measured with ICP-MS
to be 2$\times$10$^{\rm -12}$~g/g, sufficient for our requirements.
This contamination level is consistent with 
that of KamLAND-Zen 400 film, and the secular equilibrium between 
$^{238}$U and $^{214}$Bi was confirmed by $^{214}$Po($\alpha$) 
energy spectrum shape analysis in KamLAND-Zen 400~\cite{AGandoDoctorThesis}.

We can tag $^{\rm 214}$Bi decays in the film by detecting the 
$\alpha$-ray from the subsequent decay of $^{214}$Po~($\tau$ =
237~$\mu$sec) under certain conditions, namely when the $\alpha$
particle leaves the film and deposits energy in the LS, producing
a visible energy level above threshold (Figure~\ref{fig:AlphaInFilm}(b)).
A 25~$\mu$m-thick film realizes 50\% $^{214}$Po detection in KamLAND.
Although this $^{\rm 214}$Bi rejection power is limited,  this detection
can be used for background estimation. 
Thus a 25~$\mu$m thickness was preferred.

\subsection{Transparency}

A low-transparency film decreases the detector energy resolution.
This leads to an increased background level by shifting the signal from two neutrino 
double-beta ($2\nu\beta\beta$) decays into the $0\nu\beta\beta$ 
decay signal region.  
Transparency corrections could also result in a significant difference 
in visible energy scales inside and outside the IB, 
resulting in an energy distortion for events occurring near or traversing the balloon boundary. 
To avoid these problems, the transparency
should be greater than 95\% at the effective wavelength (370--400~nm).
The nylon film transparency in LS was measured to be more than 99\%. 
This level preserves the visible energy balance between energy
deposited inside and outside the IB.

\subsection{Mechanical strength}

For an IB radius of about 2~m and a density difference
between the LS inside and outside the balloon less than 0.1\%, 
we estimate a maximum stress of 4~MPa to the balloon film. 
Considering the deformation of the balloon shape, we
set a design requirement of 
40~MPa strength, allowing a safety factor of 10. 
The breaking strength of our film was measured with
a force gauge (ZTA-500N, Imada Company) to be about 
180~MPa 
for the direction of film extrusion and about 
320~MPa 
in the perpendicular direction.  
The strength of welded film is described in 
Section~\ref{sec:StrengthWeldedNylon}.

\subsection{Tightness to xenon}

\begin{figure}[htb]
\centering
\includegraphics[height=2.0in]{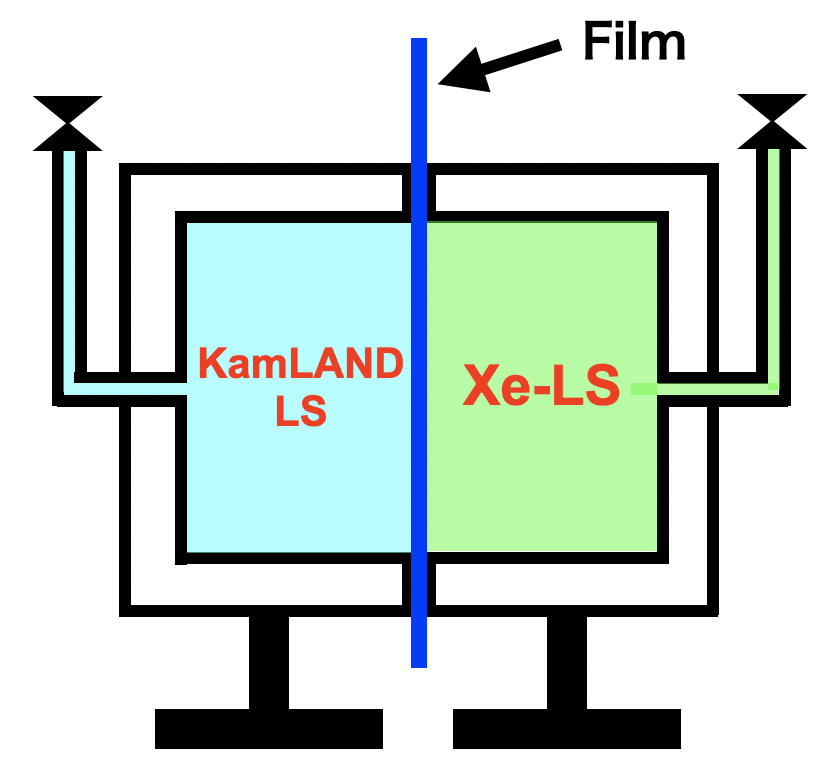}
\caption{Schematic view of xenon tightness measurement.}
\label{fig:GasTightness}
\end{figure}

We checked the tightness to xenon of the nylon film using a separated  
stainless-steel box as shown in Figure~\ref{fig:GasTightness}.  
We filled one side of the box with Xe-LS and the other side 
with KamLAND LS. 
We allowed several months to test the penetration of xenon from 
the Xe-LS to the KamLAND LS through the nylon film. 
The level of xenon in the KamLAND LS was then measured by gas chromatography 
(GC-4000 Plus, GL Sciences Company). 
We did not measure any xenon in this sample of KamLAND LS.
The estimated upper limit on the xenon leak rate from the Xe-LS 
through the IB film is 
1.9~kg over 5~years. 
This result is an acceptable level for the xenon loss and 
its impact is negligible in the systematic error for 
0$\nu\beta\beta$ decay search analysis. 

\section{Inner Balloon Fabrication} \label{sec:fabrication}

\subsection{Clean-room}

The IB was fabricated in Class 1 (ISO 14644-1) clean-rooms (super clean-room, SCR) 
at the Micro System Integration Center at Tohoku University in Sendai, 
Japan, about 600~km from the Kamioka detector site.  
Outside air is passed through multiple HEPA filters into the building.  
The entire ceiling area of each SCR is comprised of ULPA filters to
maintain a Class 1 environment 
where the clean air is flowing down from the ceiling to the grating floor.  
The size of each room is 3.8~m in height, 14.0~m in-depth, and 3.6~m in width.
We used two SCRs, 
a preparation room and a fabrication room for the IB. 
To avoid the contamination of the fabrication room, 
the preparation room was used for multiple purposes including 
welding tests, 
operational checks of devices, tool cleaning, 
and changing from normal clean-room suits to SCR suits. 
The ceiling, walls, floor, and all inner surfaces in
the SCR were wiped with ultra-pure water and clean
wiping cloths before using the room to avoid particles 
originating from these surfaces to accumulate 
on the IB film by static electricity.  
We used SFIDA and MAX-3003 (MCC Company) wiping cloths in the SCRs. 
Both cloths were packed at Class-1000 clean-room 
(FED-STD-209, consistent with Class 5 in ISO 14644-1), 
so we washed each cloth with ultra-pure water multiple times (3--5 times).  
We analyzed the particle flow from human workers, wiping cloths,
and other sources and monitored the air flow routes on the work table 
using particle visualization systems with a green laser.  
Based on these analyses, we established a procedure requiring that all 
materials, including the cloths, had to be washed with ultra-pure water, 
and that SCR wear had to be sent for cleaning after each use.

The custom-made nylon film we used for the IB reaches a static electric potential
of about +3~kV because it was fabricated without filler.  
It is therefore susceptible to contamination by dust attraction and accumulation. 
In general, humidity levels above 65\% prevent static electricity, 
but the SCR humidity is normally less than 65\% between
autumn and spring in Japan.  
We therefore set up a mist generation system just before 
the ULPA filter sending clean air to the SCR.  
The AirAKI\textsuperscript{\textregistered} (H.IKEUCHI Company) 
uses ultra-pure water and compressed air to generate 
10-$\mu$m water mist at up to 144~L/hour to maintain 
65\% humidity in the room. 
We also introduced two devices to avoid static electricity. 
A Keyence SJ-F5500 neutralization system was used on all films 
while handling the film. 
Another Keyence SJ-H neutralization bar was used when 
we took nylon films from the roll. 

\subsection{Tool cleaning}

\begin{figure}[htb]
\centering
\includegraphics[height=2.5in]{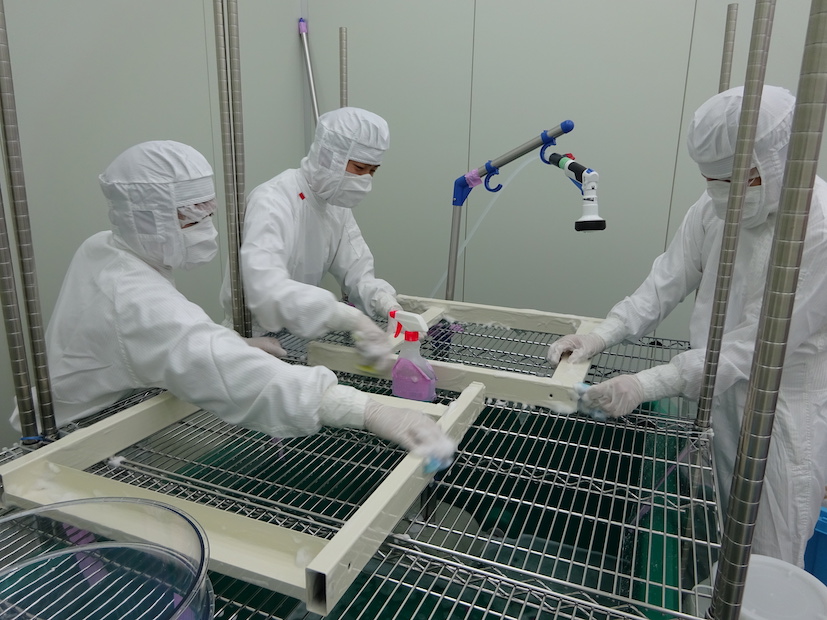}
\caption{Washing with detergent and ultra-pure water in the preparation
        room.}
\label{fig:ToolsWashing}
\end{figure}

We washed all the tools, devices, and other materials used in the 
SCR by detergent and ultra-pure water, ethanol,
isopropyl alcohol, or ethanol with ultra-sonic cleaning.  
The ethanol was mainly used to remove machine oil from the metal tools.
Isopropyl alcohol was used to clean the metal tools that touched the
IB and to remove the guide lines marked with pen on the
nylon film.  
The detergent used was Magiclean (Kao Company); this was 
employed on almost all tools and devices as shown in
Figure~\ref{fig:ToolsWashing}. 
An average of one to two tonnes of ultra-pure water 
were used per day during the washing. 

\subsection{Corrugated-tube cleaning}

The surface of the corrugated tube, which presents grooves at regular 
intervals of around 1~cm, was contaminated by large and 
small particles and by stains of machine and other oils.
The outside of the tube was
cleaned with a brush and detergent, by wiping with ethanol using a
cloth, and by ultra-sonic cleaning with ultra-pure water.  The inside
was cleaned with a brush and detergent, by flushing it with ethanol,
and by ultra-sonic cleaning with ultra-pure water. 
An abridged version of this cleaning procedure had already been applied 
to the same corrugated tube for the preparation of the KamLAND-Zen 400 
experiment: this was shown to yield sufficient cleanliness following 
in-situ measurements in the KamLAND detector.

\subsection{Film washing, cover setting and cutting}

\begin{figure}[htb]
\centering
\includegraphics[height=2.1in]{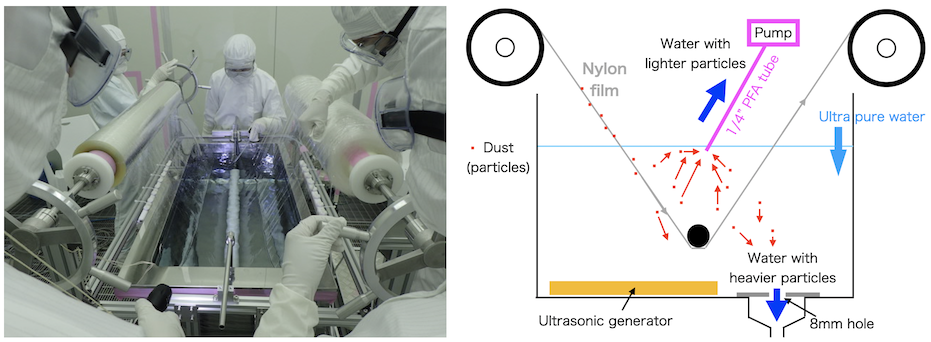}
\caption{Ultra-sonic film cleaning in ultra-pure water.  The film
        speed was controlled with two rudders by hands.  The 
        white center bar is used to submerge the film.  Dust in the
        water was monitored with LED light and removed with a pump.
        A neutralization bar was set under the film roll to prevent
        the film from sticking due to static electricity.
        The ultra-pure water was fed continuously during  film washing, 
        and particle contaminant was removed by a pump 
        at the water's surface and by a draining line 
        at the bottom of the tub.}
\label{fig:FilmWashing}
\end{figure}

The first step of IB fabrication was cleaning the nylon
film to remove surface contamination.  We used ultra-sonic cleaning
in ultra-pure water in a stainless-steel tub as shown in
Figure~\ref{fig:FilmWashing}.  
The film speed was set to about 40 m/hour, 
a balance between the need to drain the water off 
the nylon by gravity and minimizing the risk of water
sticking to the film due to surface tension.
Light particles were removed from the water by pumping 
the water at the surface level. 
Heavier particles 
were removed through three 8-mm diameter draining holes at the 
bottom of the cleaning tub.  
We applied this cleaning procedure twice, thus reducing 
the $^{\rm 238}$U level from 
9$\times$10$^{\rm -12}$~g/g to 2$\times$10$^{\rm -12}$~g/g, 
as measured by ICP-MS at NTT-AT Co.

We set cover films over both sides of the IB film during
the several-month period required of balloon fabrication to avoid dust
accumulation due to static electricity.  
The cover films were made of the same
clean film as the IB and adhered to the IB film 
by static electricity.  
When we cut the film into the 
balloon patterns we cut all three layers simultaneously.  
During the welding, we turned back the edges of the cover films, 
and after the balloon sections had been joined, the cover-film edges 
were unfolded to cover and protect the welding lines.

Nylon film was cut into 32 rectangles with dimensions
7.5~m $\times$ 1~m to make the 24 gore sections, 6 cone 
sections, the polar cap and polar harness, and one straight tube section, 
as shown in Figure~\ref{fig:mini-balloon}. 
After cutting we made alignment marks on the film for welding
position checks.  These marks were made with a Microperm 03 pen
(Sakura Color Products Company).  
The $^{\rm 238}$U level of the ink was measured by ICP-MS at 
NTT-AT Co.~to be 1.2$\times$10$^{\rm -10}$~g/g, 
an acceptable value considering the tiny amount of 
remnant ink on the IB.

\subsection{Welding}
\label{sec:StrengthWeldedNylon}

Nylon films are typically joined with glue.  The Borexino
experiment~\cite{BOREXINO} used resorcinol to glue sections of
125-$\mu$m nylon film together.  However, the glues we tested made
many holes in 25-$\mu$m nylon film, and we could not find suitable
gluing procedures to avoid this problem.  Working with DAIZO
Corporation, we instead developed a method to join nylon films by
heat welding.  The melting point of Nylon-6 is about 225 $^\circ$C. 
The components of our film are 30\% crystallized nylon
and 70\% amorphous nylon~\cite{ItoSanTOYOBO}.  As a result, 
low-temperature welding at less than 225 $^\circ$C is possible by
molecular motion.  However, in our welding tests we found that
high-temperature welds were stronger than low-temperature welds.
We therefore tuned the welding parameters based on high-temperature
welding.

\begin{figure}[htb]
\centering
\includegraphics[height=4.5in]{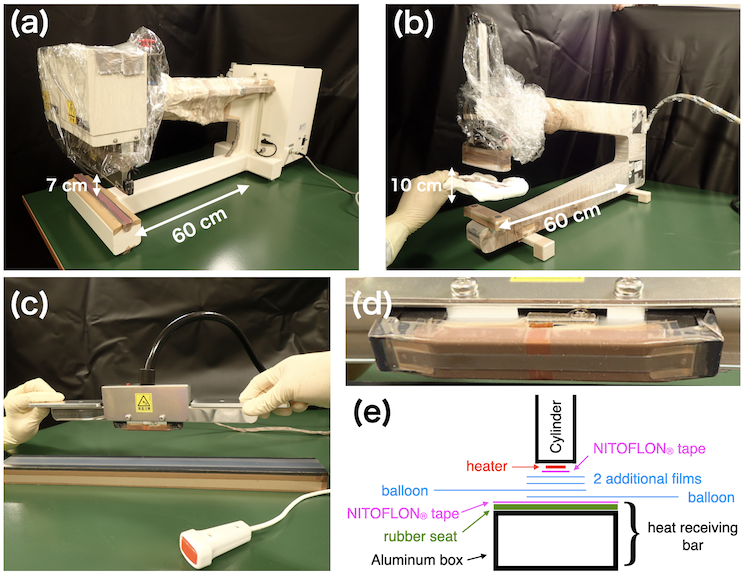}
\caption{(a) Primary welding machine.
        (b) Secondary welding machine.  The lower heat-receiving
        bar can be separated and set on top of a different section
        of the IB in order to weld the final seams.
        (c) Hand welding machine. 
        (d) The tape-covered heater of the hand welding machine.
        (e) Schematic of film welding.  The four film layers are 
        compressed by the cylinder and heat applied.  Layers of
        NITOFLON\textsuperscript{\textregistered} tape under the 
        heater and above the lower rubber
        seat prevent melted film sticking to the welding device.
        The heat-receiving bar is used to reflect heat from the
        heater.}
\label{fig:Weldingmachine}
\end{figure}

\begin{figure}[htb]
\centering
\includegraphics[width=0.95\linewidth]{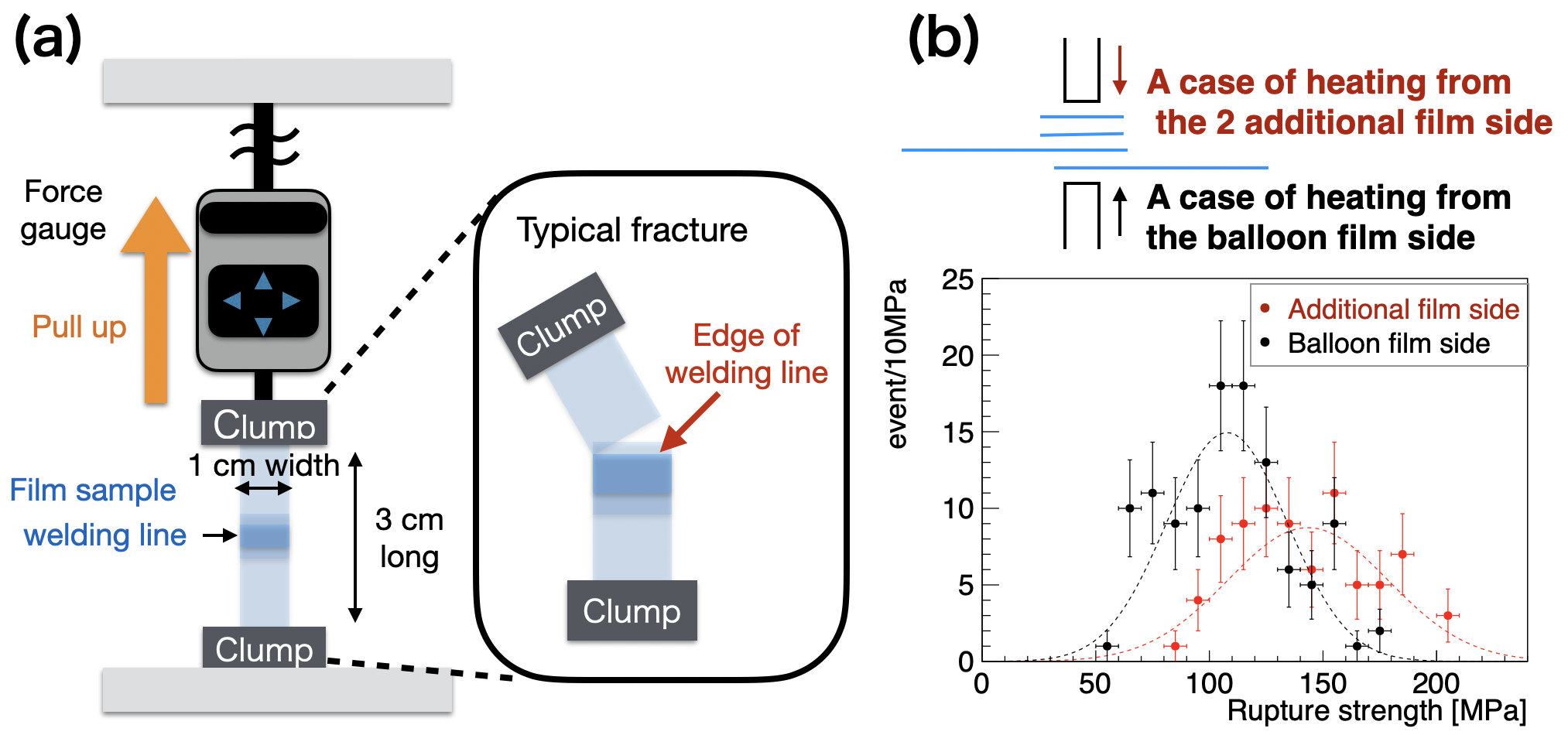}
\caption{(a)  Schematic view of the weld strength test.   Fractures
        typically occur at the edge of the weld.
        (b)~Film strength after welding.  The data in red are from welds
        made with heat applied through the two additional layers 
        of film, while black shows data for welds made with heat applied directly to the IB material.}
\label{fig:WeldingStrength}
\end{figure}

The primary welding machine was made by Fuji impulse Company
(Figure~\ref{fig:Weldingmachine}(a)).  
It generated impulse heating over a 5~mm $\times$ 
310~mm area and could apply heat between 180 to
250~$^\circ$C for 0.5 to 10~seconds. 
NITOFLON\textsuperscript{\textregistered} tape 
(Polytetrafluoroethylene base tape, Nitto Denko Corporation) 
was applied over the heater surfaces to prevent the melted 
film from sticking to the device. 
We found that a two-layer welding with the primary welding machine 
produced many holes in the weld.
To avoid this problem, we added two additional layers of film along the seams, 
for a total of four layers with a combined thickness of 100~$\mu$m. 
A schematic of the final welding configuration is shown in
Figure~\ref{fig:Weldingmachine}(e).
The additional films melt under
heating and filled in holes in the IB film.  They also
avoided overheating the IB film.  After testing a range
of parameters, we established a welding procedure using 225~$^\circ$C
heating for 3.5~seconds.  
The pressing pressure was released when
the temperature reached 80~$^\circ$C.  The strength of the welded seams
was measured with a force gauge as shown in
Figure~\ref{fig:WeldingStrength}(a).  Figure~\ref{fig:WeldingStrength}(b)
shows the film strength after welding.  This figure indicates that
heating through the added films produces a stronger weld than
heating the IB film directly. 
During balloon production, all seams were welded by applying heat through the added films.

Some of the welds required to form the IB presented
special challenges.  To make the final weld between the gores forming
the main body of the sphere, the upper heater of the welder had to 
be outside the volume of the balloon while the heat-receiving bar
had to be inside, a topological impossibility with the primary welding
machine.  In several regions where weld lines converged we needed
shorter welds than the 31-cm lengths of the primary welding machine,
namely in the areas near the bottom pole, at the junction between
the gores and the upper cone, and at the connection between the
upper cone at the straight tube.  We worked with Creative Design
and Engineering (Laporte, Colorado, USA) to develop a secondary
welding machine (Figure~\ref{fig:Weldingmachine}(b)) to address
these issues.  The secondary machine features a 10-cm heater and a
lower heat-receiving bar designed to allow the second layer of film to
pass under the bar.  The best welding parameters for the secondary
machine were welding at $\sim$225 $^\circ$C for 10.0~seconds.  Welds
made with the secondary machine had strength greater than 
160~MPa, slightly stronger than the primary machine.

The welding parameters of the primary and secondary machines were tuned 
to provide the nominal strength. 
However, welding still periodically resulted in holes along the weld lines. 
The number of holes in the welds was
variable and depended on additional factors that we could not completely
control, including the room temperature, the humidity, and the
condition of the NITOFLON\textsuperscript{\textregistered} 
tape between the IB film and the heater.  
We therefore developed a methods 
to patch these holes, as is described in Section~\ref{sec:LeakCheckRepair}.

Most of the welding lines were made in 30-cm increments using 
the primary welding machine.  
However, since the heat-receiving bar opposite the heating element 
in this machine could not be detached, the final weld line in the 
main body of the sphere was kept open, as were other welds 
in areas where weld lines converged.  
These final welds were made in smaller increments with 
the secondary welding machine (Figure~\ref{fig:Weldingmachine}(b)), 
with a few particularly 
difficult areas done using a hand welding machine 
(Figure~\ref{fig:Weldingmachine}(c)).  
Both these smaller machines allow the heat-receiving bar 
to be positioned inside the balloon volume.

\subsection{Connection cylinder}
\label{sec:FilmWithCylinder}

The IB and the upper corrugated tube are connected by a
plastic cylinder made of PEEK, by using PEEK film holders, additional
nylon film, and glue, as shown in Figure~\ref{fig:CorrugatedTube}(b).
The outer circumference of the connection cylinder and the inner
surface of the top of the straight-tube section of the IB 
are a tight fit.  Once connected, they can support 20~kg.  However,
to reduce the risk of slippage of the IB, we clamped the
IB film with outer film holders made of PEEK.  
We applied glue
to fill any small gaps that might allow gas or liquid to leak.
The glue we used was ``Adcoat'' a mixture of 91\% AD-76P-1 with 9\%
CAT-10L (Toyo-Morton Company, $^{\rm 238}$U $<$1$\times$10$^{\rm -11}$~g/g measured with ICP-MS). 
This glue was selected for its suppleness based on our prior 
experience with its use as the laminate glue in the
construction of the original (outer) KamLAND balloon. 
It takes more than one day to harden, 
but it is suitable
for these types of connections, which require multiple steps for the
overall gluing operation, namely fitting the film onto the cylinder,
arranging the glue and additional film layers, and installing
the film holders, making a glue with a setting time preferable.  
We actually allowed 3~days for the glue to harden at room temperature.  

\subsection{Leak check and repair}
\label{sec:LeakCheckRepair}

\begin{figure}[htb]
\centering
\includegraphics[height=2.5in]{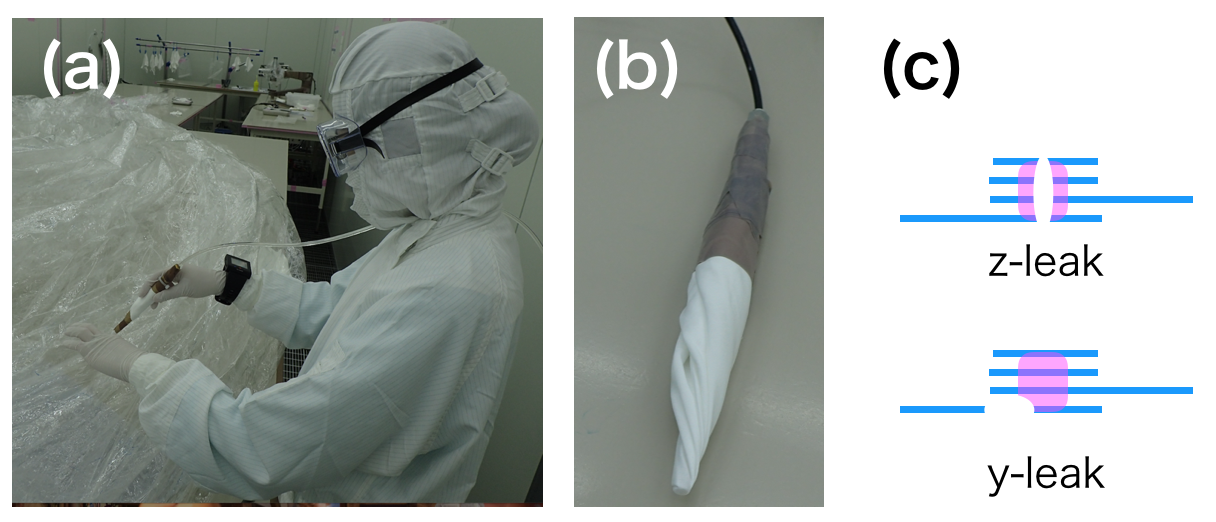}
\caption{(a) Leak check. The IB was filled with helium
        and leaks were investigated with a helium detector outside the 
        balloon.  The leak rate was viewed via a wrist monitor 
        connected to a camera in front of the helium detector.
        (b) Close-up photo of leak detector probe covered with SFIDA to avoid the damage to the 
        film or contamination with dust.
        (c) Definition of welding line leak types.  The layers of
        nylon film are shown in blue, while the pink area represents 
        the hardened nylon melted by the welding processes.  Defining
        the x-axis to lie along the weld line, a z-leak penetrates the weld in the direction normal to the film surface while a
        y-leak corresponds to a penetration escaping the side of the weld, perpendicular to both a z-leak and the weld line.}
\label{fig:LeakCheck}
\end{figure}

We checked the balloon for leak points by filling it with helium and
using helium leak-check detectors (M-222LD-D, Canon Anelva Corporation, and
Adixen, Pfeiffer Vacuum GmbH), as shown in Figure~\ref{fig:LeakCheck}(a).
To avoid introducing dust to the IB surface we covered
the helium probes with SFIDA cloth (Figure~\ref{fig:LeakCheck}(b)).
We found two types of leaks, termed z-leaks and y-leaks, as illustrated
by Figure~\ref{fig:LeakCheck}(c).  The z-leaks appear to be caused
by factors including the boiling of water contained in the film,
local heat concentration by the NITOFLON\textsuperscript{\textregistered} tape
adhering to the film, the condition of the heating element, 
creases in the film, and air
remaining between the films.  The y-leaks appear to be caused by
insufficient welding heat, and also by film creases and local heat
concentration.  The total number of leak points was about 1,000. 
Most of these holes were not visible, but they included about 10 
holes of about 0.1 mm width and several y-leaks of about 10 mm width. 

Two methods were used to repair the leak holes made during film welding,
both based on the use of additional nylon film, i.e. 
re-welding and gluing.  The largest
leak points, mainly torn parts along welding lines, were repaired
with re-welding.  However, after repeated re-welds the nylon film
becomes harder, and re-welds create weak points susceptible to 
damage when the film is folded.  Therefore most leaks were repaired
with glue and small pieces of nylon film (10~mm $\times$ 10-30~mm).  
The glue used was Aron Alpha 202 (Toagosei Company, Ltd.).  
This glue is a single-liquid adhesive that cures at room temperature 
and humidity, and has high viscosity.  
It did not adhere to hardened glue, so it could not
be used as the primary film connection method, but its properties
were suitable for repairing the many small leaks.  
The glue passed compatibility tests in LS, and ICP-MS measurements 
at NTT-AT Co.~gave an upper limit on the $^{\rm 238}$U concentration of 
$<$5$\times$10$^{\rm -12}$~g/g.
The surface of the glue hardened in several seconds, 
but we allowed more than 12~hours for the full area to solidify.  
The repaired points were checked with the helium leak detector 
after the glue hardened.

\subsection{Hanging belts for the inner balloon}
\label{sec:HangingBelt}

\begin{figure}[htb]
\centering
\includegraphics[width=0.95\linewidth]{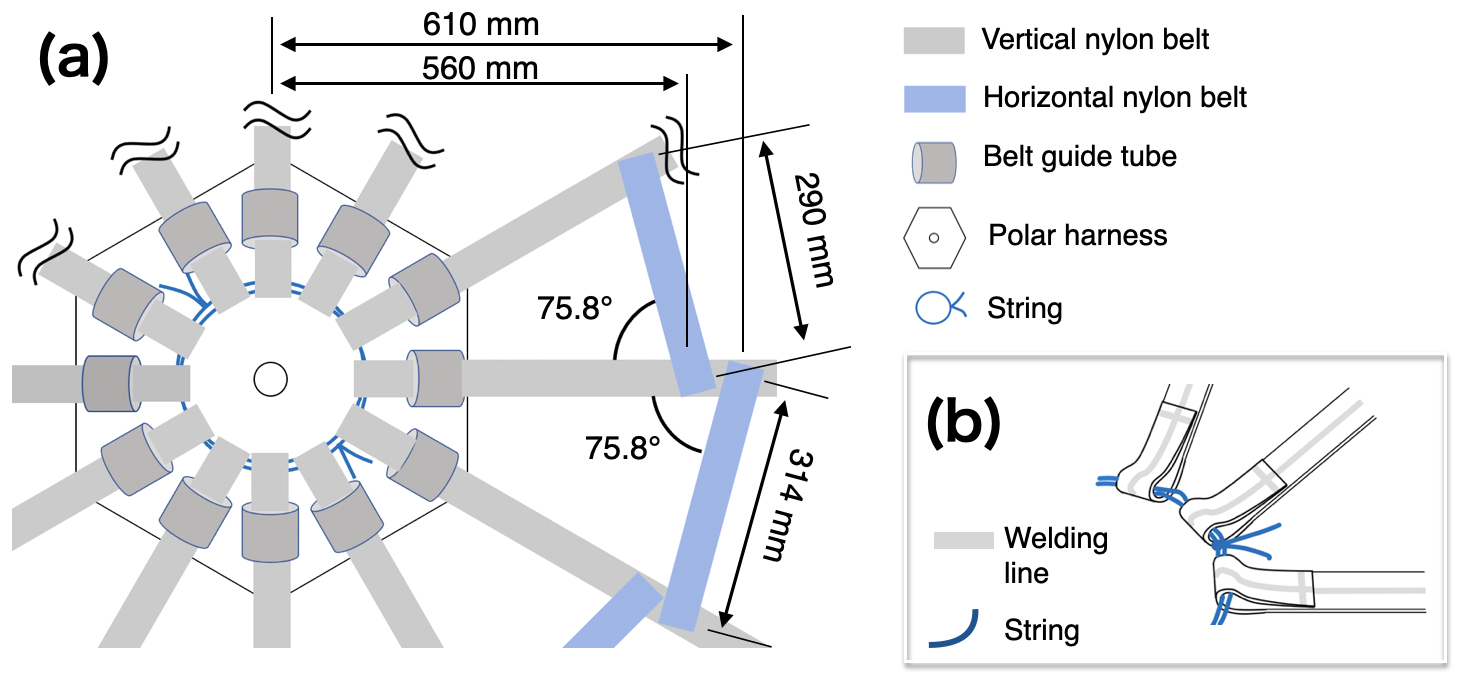}
\caption{
        (a) Schematic view of the polar area detailing the hanging belts for the IB. 
        The connection angle of the two belts is designed to be parallel 
        to the equator when the balloon is inflated.
        (b) Schematic view of the connections between the ends of the 12 vertical belts 
        with the 2 Vectran strings.
        }
\label{fig:Belt}
\end{figure}

The bottom of the IB rests on the polar harness, which is connected to the
12 load-bearing nylon hanging belts as shown in Figure~\ref{fig:mini-balloon}. 
Figure~\ref{fig:Belt}(a) shows a schematic view of the polar 
area including the belt connections. 
The polar harness, which has a vent hole at the center, was made 
from a single layer of nylon film. 
The purpose of the vent hole was to avoid air getting trapped between 
the harness and the polar cap during installation, which would lead 
to light reflection that could bias vertex reconstruction.
The belts were made of 2 layers of nylon film welded along the center-line and at their ends.
To avoid damage from local tensions, 
the hanging belts were not directly welded to the polar harness film. 
The position of the vertical belt is constrained by guide tubes 
that are also made of nylon film and are glued to the polar harness with Aron Alpha 202. 
The ends of the 12 belts were welded into loops and connected with 2 Vectran 
strings~(Figure~\ref{fig:Belt}(b)). 
We also connected horizontal belts to the vertical belts 
near the polar harness with Aron Alpha 202 
to constrain the relative position of the belts. 
The horizontal belts are similar in structure to the vertical belts 
but their ends were not closed by welding.
The position of the belt and harness assembly relative to the polar cap was left unconstrained, as full-size IB installation tests performed for the KamLAND-Zen 400 balloon demonstrated successful positioning of the polar harness directly below the polar cap in four out of four attempts.
The belt and harness assembly was set to the IB just before packing and delivery to the detector site.

\subsection{Packing}

\begin{figure}[htb]
\centering
\includegraphics[height=1.7in]{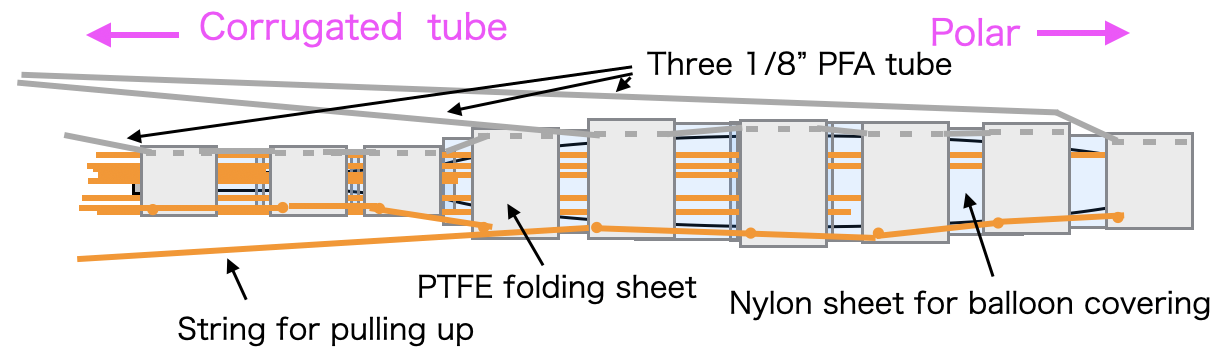}
\caption{Schematic view of how the IB was folded and
        packed.  The folded balloon was covered with a nylon sheet, 
        held in place by sheets of PTFE.  
        The PTFE sheets were
         closed by 1/8'' PFA 
        tubes, released by pulling out
        the tubes.  The strings attached to the PTFE 
        and nylon cover sheets allowed them to be extracted from KamLAND
        following the IB insertion.}
\label{fig:Foldings}
\end{figure}

The interior of KamLAND can only be accessed via a narrow
opening through the top ``chimney'' region of the detector  
(see Figure~\ref{fig:KamLAND}).
To pass through this space we folded the IB to 
within a 50-cm diameter.  
To protect it from damage and to avoid flexing, especially
in the small clean-room above the chimney, the IB was
folded and packed inside commercial nylon and PTFE 
sheets as shown in Figure~\ref{fig:Foldings}.  
The PTFE sheets were closed by 1/8'' PFA 
(perfluoroalkoxy alkanes) tubes, 
released by pulling out the tubes. 
The strings tied to PTFE and nylon sheets were used to 
pull up sheets from the detector after sinking the 
IB (Section~\ref{sec:install}).
A custom plastic delivery box was made by
the Naigai TEC Corporation.  The inner surface was lined with film
made by Daizo Corporation (KS137-205).  KS137-205 is a 205-$\mu$m
film made of aluminum vapor deposition EVOH with low density
polyethylene on both surfaces for gas tightness.  
After packing the IB inside, the box was set inside 
a 1.8~m~$\times$~1.8~m~$\times$~0.7~m bag made of KS137-205, 
and it was sealed via welding. 
This airtight bag kept the balloon in the SCR air and humidity.

\section{Installation in the Detector} \label{sec:installation}

\subsection{Onsite preparation}

In order to insert the IB inside the OB, a clean tent was built 
on the top of the KamLAND detector.
The clean-room class grade was between 10--100 (FED-STD-209) 
depending on the area and the number of people in the tent.  
The inner surface of the room was cleaned with lint rollers 
and by wiping with ultra-pure water.  
The ultra-pure water, with electrical resistivity 
greater than 18.2~M$\Omega\cdot$cm, 
was made using reverse osmosis apparatus with a RFU554CA
(Advantec MFS, Inc.) filter.  
It was used for wiping down surfaces, washing gloved hands, 
and keeping tools clean.  
The tables, tools, wipers, and other materials were all
delivered to the KamLAND site from Sendai after being cleaned 
and packed in the SCR.

\subsection{Corrugated tube connection}

A specific procedure was developed and set up in order to 
assemble the corrugated tube. 
The individual steps consisted in:
\begin{enumerate}
    \item Set two O-rings in adjacent ``valley'' near the end of 
            the corrugated tube (see Figure~\ref{fig:CorrugatedTube}).
    \item Insert the corrugated tube into the PEEK connection cylinder.
    \item Leak check the connection.
    \item Apply Adcoat glue at the edge of the PEEK cylinder.
    \item Wait 5~days for the adhesive to harden.
    \item Connect the grip piece.
\end{enumerate}
We used two O-rings for each connection because we found that the 
deformation of the corrugated tube led to leaks when only one was used. 
We added glue to ensure leak tightness
as the O-rings alone may leak (Xe-)LS or gases.

\subsection{String guide}

The string guide (Figure~\ref{fig:mini-balloon})
drives the 12 strings,
preventing them from touching the outer balloon or the chimney of the
detector, and avoiding any entanglement of the strings themselves.
The string guide is positioned under a light-blocking black film
at the bottom of the chimney.  
We located it as high as possible to prevent the nylon support 
belts from cutting into the IB.  
The string guide is connected to the corrugated tube. 
It can be rotated relatively to the fixed corrugated tube to 
remove any twisting of the balloon or belts and ropes 
caused by the installation processes.

\subsection{Monitoring camera}

\begin{figure}[htb]
\centering
\includegraphics[height=1.95in]{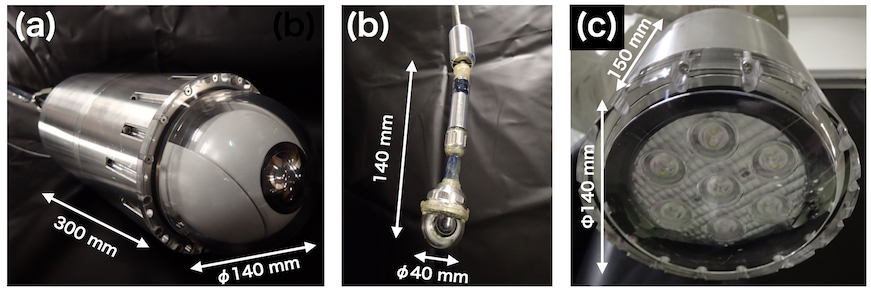}
\caption{(a) Primary monitor camera.  The IP camera is encased
        in a stainless steel and acrylic cover for use in the LS.  
        Power and data are both transferred via a PoE cable encased
        in PFA that runs to the top of the detector.  
        (b) Secondary camera.  The camera cover and cable were
        replaced by acrylic and a fluorine-surface cable, respectively.
        (c) Lighting for detector interior.  LED lights (2,200~lm) 
        with encasement similar to that of the primary camera.}
\label{fig:camera}
\end{figure}

We cannot view the interior of the detector directly through the
upper opening because the hole area at the bottom of chimney is
covered by black sheets to block light from the outside.  
We used 
two monitoring cameras compatible with LS to oversee 
the IB installation and to ensure the
safety of both the IB and the rest of the detector.  
After the installation, the cameras were used to regularly monitor 
the condition of the detector and IB for several months. 
The primary camera, shown in  
Figure~\ref{fig:camera}(a), is a high-resolution Internet Protocol (IP) camera that can
resolve small details and view the whole IB and detector.
This camera was covered by a stainless-steel housing and a UV-transparent
acrylic dome for use in LS.  The power supply and
data transfer are through a Power over Ethernet (PoE) LAN cable encased in PFA tubing.
The primary monitoring camera was installed before the IB
installation and was used to monitor the IB during installation,
to monitor balloon expansion during LS filling, to search for signs
of leakage, and to assist the positioning and retracting of the
filling/draining tube.  Figure~\ref{fig:camera}(b) shows the secondary
camera, based on a camera designed for pipe inspection (W\"{o}hler
VIS 350, W\"{o}hler Technik GmbH).  This camera was used to check
the vicinity of the IB.  The polycarbonate cover dome
of this camera was replaced with UV-transparent acrylic for use in
the LS.  The fiberglass cable was also replaced with one with a
fluorine surface compatible with the LS.  We also positioned LED
lights in the detector, illuminated only while actively monitoring the IB (Figure~\ref{fig:camera}(c)).

\subsection{Installation}
\label{sec:install}
  
\begin{figure}[htb]
\centering
\includegraphics[height=2.5in]{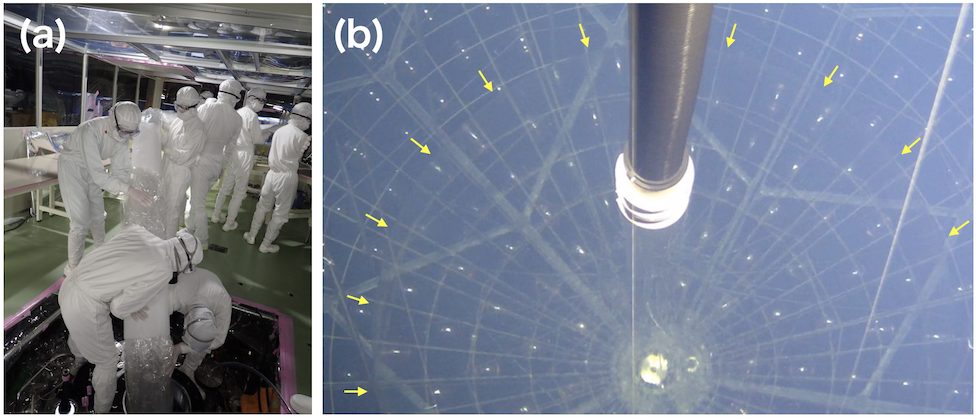}
\caption{(a) Installation of the IB into the detector.
        (b) Photograph of the IB installed in the detector, 
        taken with the monitoring camera.  It is difficult to see
        the IB directly; the edge shown by arrows can be located 
        by noting the discontinuities in the imaged stainless
        steel structures.}
\label{fig:IBInstallation}
\end{figure}

The densities of nylon and PTFE are 1.2~g/cm$^{\rm 3}$ and
2.2~g/cm$^{\rm 3}$, respectively, while that of the KamLAND LS is 
0.77728~$\pm$~0.00004~g/cm$^{\rm 3}$. 
Therefore, the IB with its folding covers would ideally 
sink into the LS.  
However, due to air remaining in the IB, lowering it into
the detector is actually not so easy.  
We introduced pseudocumene-rich LS with a density of about 
0.78~g/cm$^{\rm 3}$ into the IB
to extract the residual air and to decrease the balloon's buoyancy. 
As we lowered the IB, we repeated the step-by-step 
filling in 5--10~L increments.  
We lowered the balloon as shown in
Figure~\ref{fig:IBInstallation}, keeping the balloon folded in its
nylon and PTFE covers (Figure~\ref{fig:Foldings}).  
Once this was in position, we removed the 1/8'' PFA tubes 
that secured the PTFE cover sheets so that the sheets 
would detach from the balloon.  
We then used the strings attached to the nylon and PTFE covers 
to pull them up to the access area at the top of the detector 
and remove them.

\subsection{Inner balloon inflation and liquid scintillator filling}

The IB was inflated by filling it with decane and pseudocumene based LS 
(non-Xe-loaded LS) with a density 0.77732~g/cm$^{\rm 3}$.  
Since the density of LS increases by 0.1\%/$^\circ$C at temperatures 
near 10$^\circ$C, we maintained the temperature of the added scintillator
within the range 9.5--10.5$^\circ$C, compared to a temperature at
the center of KamLAND of about 12.0$^\circ$C.  The LS flow rate was
about 270~L/hour, and the total volume of LS filled to expand the 
IB was 30.5~m$^{\rm 3}$.
The effective radius of the IB 
based on the filled non-Xe-loaded LS is 1.90~m.

\subsection{Filling of xenon loaded liquid scintillator}

\begin{figure}[htb]
\centering
\includegraphics[height=6.6in]{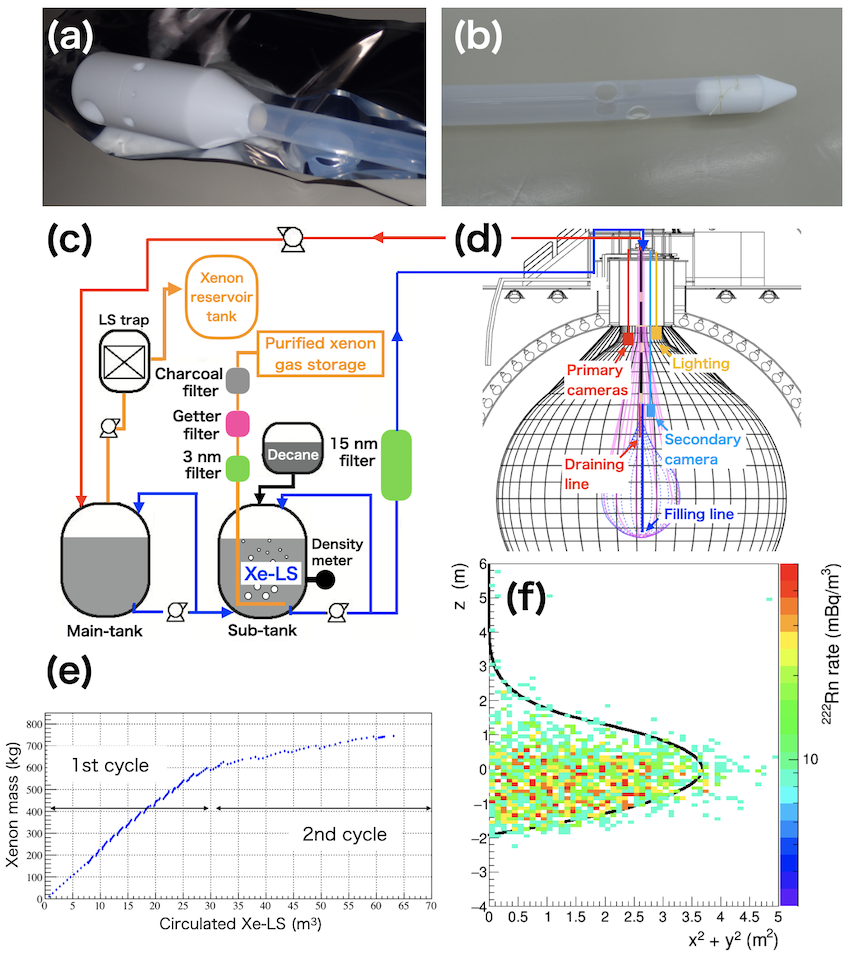}
\caption{(a) The end of Xe-LS filling line. 
        (b) The end of LS draining line. 
        (c) Diagram of the Xe-loading circulation loop, 
        replacing LS with Xe-LS. 
        (d) The setup positions of cameras and tubes in the detector. 
        Two primary cameras placed diagonally across the corrugated tube 
        monitor the entire balloon. 
        (e) Xenon mass v.s. circulated LS volume. 
        (f) The vertex distribution of $^{\rm 214}$Po events during Xe-LS filling. 
        The black line is the expected outline of the IB film position.}
\label{fig:XeLS}
\end{figure}

The non-Xe-loaded LS was replaced by Xe-LS by circulation 
after the IB was leak-checked.
The leak checked was performed by monitoring the IB weight with load cells, 
the shape stability was verified with the primary cameras 
and the decane concentration in the KamLAND-LS was measured using gas chromography. 
A 3/4'' PFA tube was used as a filling line (Figure~\ref{fig:XeLS}(a)), 
with its end positioned 200~mm above the bottom of the balloon 
(Figure~\ref{fig:XeLS}(d)). 
A similar tube, whose end is positioned in the cone area, was used as  
draining line (Figure~\ref{fig:XeLS}(b)(d)). 
The xenon was dissolved in the non-Xe-loaded LS removed 
from the IB by bubbling xenon in a dedicated circulation system 
(Xe-system, Figure~\ref{fig:XeLS}(c)).
The non-Xe-loaded LS was first degassed in the main tank. 
Both xenon and decane were then added to the LS in the sub-tank 
to control the density of the Xe-LS. 
The Xe-LS was sampled from the sub-tank flow line and the xenon 
concentration was measured by gas chromatography before 
filling to the IB. 
The Xe-LS was tuned to have a density of about 0.77741~g/cm$^{\rm 3}$ 
and temperature of about 10.5$^\circ$C. 
This Xe-LS had a slightly higher density than the non-Xe-loaded LS in the IB, 
and this allowed to keep the non-Xe-loaded LS separated 
from the Xe-LS in the IB.
The Xe-LS volume was monitored by $^{\rm 214}$Bi--$^{\rm 214}$Po 
coincident events of $^{\rm 222}$Rn daughter nuclei 
as shown in Figure~\ref{fig:XeLS}(f). 
The source of the $^{\rm 222}$Rn, whose half-life is 3.8~days, 
is emanation from the inner surfaces of the stainless steel tanks 
in the Xe-system, and from the piping between 
the Xe-system and the PFA tubes at the detector. 
The Xe-LS and non-Xe-loaded LS flow rate was about 150~L/hour, 
with about 600~L in one filling batch due to the size of sub-tank 
in the first cycle. 
We performed roughly two full volume exchanges of the Xe-LS to 
maximize the xenon concentration (Figure~\ref{fig:XeLS}(e)), 
as some amount of the Xe-LS mixes with the non-Xe-loaded LS at the boundary. 
We applied continuous Xe-LS circulation between the sub-tank and the IB 
in the second cycle to maximize the xenon concentration in Xe-LS.
The total amount of xenon dissolved into the balloon
was measured by the weight of gas storage to be
745~kg, corresponding to 677~kg of $^{\rm 136}$Xe. 
This is consistent with the estimate from the Xe-LS xenon 
concentration analysis by gas chromatography. 

\section{Inner Balloon Contamination}\label{sec:analysis}

\begin{figure}[htb]
\centering
\includegraphics[height=2.5in]{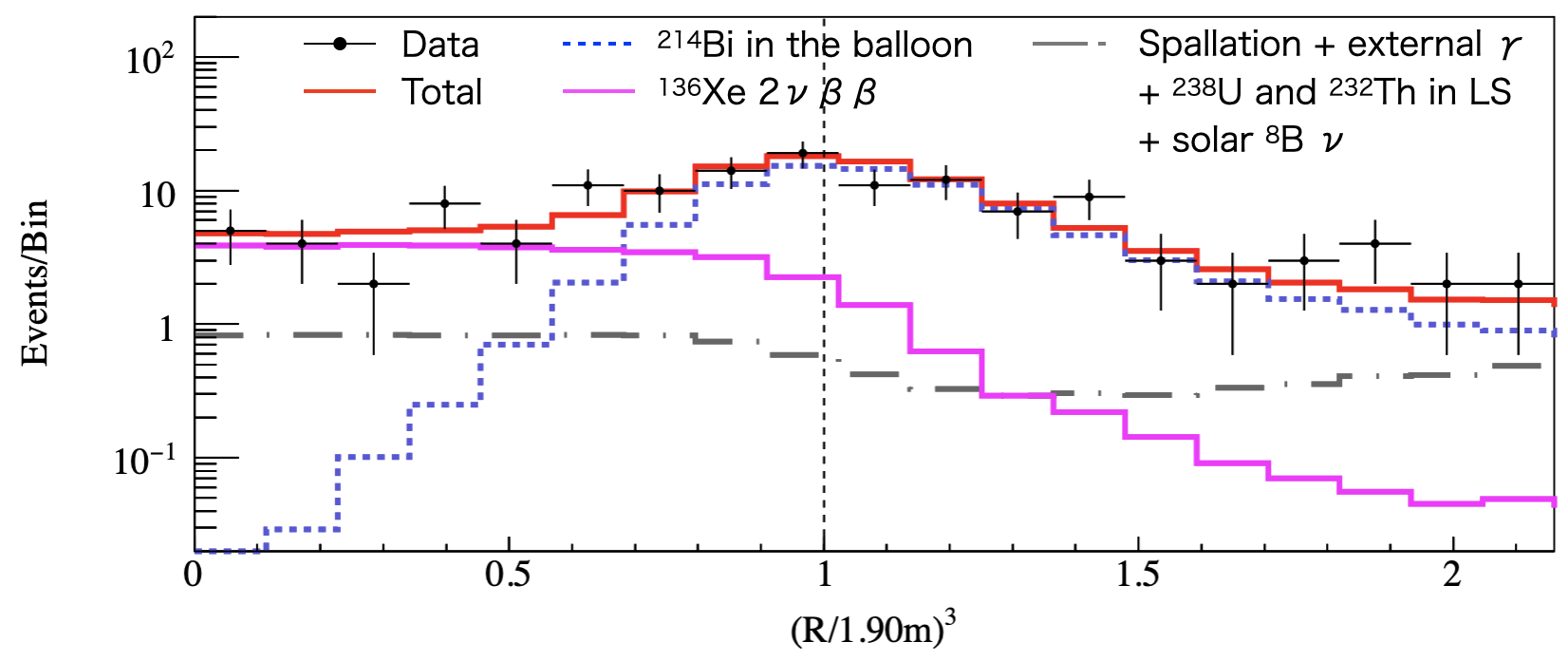}
\caption{Reconstructed radial distribution of events with visible energy 2.30--2.70~MeV
        after filling the IB with Xe-LS.  
        The radial position is normalized to the nominal IB radius of 1.90~m.
        The observed data are plotted in black with error bars.
        The red line is the total fit to the data, 
        while the blue dotted line is the best-fit contribution from $^{\rm 214}$Bi in the balloon film, 
        the magenta curve is $2\nu\beta\beta$ decay, and the gray dashdotted line represents the sum of all other backgrounds: 
        muon spallation products of carbon and xenon, 
        external $\gamma$, $^{\rm 238}$U and $^{\rm 232}$Th-series
        in LS, 
        and solar $^{\rm 8}$B neutrinos.
}
\label{fig:FilmAnalysis}
\end{figure}

The levels of $^{\rm 238}$U and $^{\rm 232}$Th in and on the deployed 
IB film were assayed in-situ by measuring the excess spectral 
strength due to these sources reconstructing at the radial position 
of the balloon surface.  
Figure~\ref{fig:FilmAnalysis} shows the radial distribution of events 
with visible energy range 2.30~MeV~to~2.70~MeV, 
which is dominated by $^{\rm 214}$Bi decay events.
We estimated the number of events of $^{\rm 214}$Bi and 
$2\nu\beta\beta$ by fitting the observed data simultaneously 
for both components. 
The other backgrounds such as  muon spallation products, 
external $\gamma$ from 
the stainless steel containment vessel, 
$^{\rm 238}$U and $^{\rm 232}$Th-series in LS, 
and solar $^{\rm 8}$B neutrinos 
were estimated using the KamLAND LS volume and fixed in the fitting. 
From this analysis and that of $^{\rm 208}$Tl, we evaluated 
$^{\rm 238}$U and $^{\rm 232}$Th levels in the IB as
3$\times$10$^{\rm -12}$~g/g and 4$\times$10$^{\rm -11}$~g/g, 
respectively.  
Contamination was reduced by a factor of 10 from the KamLAND-Zen~400 
phase of our $0\nu\beta\beta$ search~\cite{zen400final}.
The amount of $^{\rm 238}$U contamination is consistent 
with pre-installation measurements of the raw nylon film, 
thus proving that we succeeded in constructing a new IB 
without polluting the detector.

Looking ahead, to reach even lower
background levels due to radioactive
contamination of the IB, we would need to either use cleaner
nylon film, produced in a clean-room, or to use a different technique
like introducing a self-vetoing balloon made from scintillating material~\cite{scinti-balloon}.

\section{Summary}

We fabricated a balloon to contain the xenon-loaded liquid scintillator
for the KamLAND-Zen 800 experiment. 
The balloon was made of nylon film and assembled in an ultra-clean environment.  
We successfully installed the balloon and inflated it inside the KamLAND detector 
by filling it with density tuned liquid scintillator without xenon. 
Finally, we successfully replaced the liquid scintillator with 
the xenon-loaded liquid scintillator.
The $^{\rm 238}$U contamination of the balloon is consistent with
the level in the initial nylon, indicating that we built the balloon
without significant additional contamination.  
This contamination level shows a factor-10 improvement with respect to KamLAND-Zen~400.


\acknowledgments

The authors would like to thank the DAIZO Corporation for 
developing nylon film welding with us, Toyobo Company for 
developing and providing the custom-made nylon film, 
and Creative Design and Engineering for developing 
secondary welding machine.   
We are grateful to the institutes and companies 
described in the main text, all of whom have been 
critical in the production of the IB and 
the development of associated technologies. 
The KamLAND-Zen experiment is supported by JSPS KAKENHI Grants 
No. 21000001 and No. 26104002; the World Premier International 
Research Center Initiative (WPI Initiative), MEXT, Japan; Stichting
FOM in the Netherlands; and under the U.S. Department of Energy
(DOE) Contract No. DE-AC02-05CH11231, 
as well as other DOE and NSF grants to individual institutions. 
The Kamioka Mining and Smelting Company has provided 
service for activities in the mine. 
We acknowledge the support of NII for SINET4.  
This work was supported by JSPS KAKENHI Grants No. 21244025, 
No. 25220704, No. 17H01120, No. 21684008, No. 26287035, and No. 18J10498, 
NSF Award 1806440 and a UVLAC-Hayashi MISTI Seed Grant.
This work was partly supported by the Graduate Program on Physics 
for the Universe (GP-PU), Tohoku University.

\end{document}